\documentclass[sigconf]{acmart}

\renewcommand\footnotetextcopyrightpermission[1]{} 
\usepackage{graphicx}
\usepackage{mathtools, nccmath}
\usepackage{longtable}
\usepackage{array}
\usepackage{multirow}
\usepackage{booktabs}
\usepackage{tikz}

\usepackage[colorinlistoftodos]{todonotes}
\usepackage[short]{optidef}

\setcopyright{none}

\usepackage{lipani_math}

\begin{document}
\title[Towards More Accountable Search Engines]{Towards More Accountable Search Engines: \\ Online Evaluation of Representation Bias}

\author{Aldo Lipani}
\affiliation{%
  \institution{University College London}
  \city{London}
  \country{United Kingdom}
}
\email{aldo.lipani@ucl.ac.uk}

\author{Florina Piroi}
\affiliation{%
  \institution{TU Wien}
  \city{Vienna}
  \country{Austria}
}
\email{florina.piroi@tuwien.ac.at}

\author{Emine Yilmaz}
\affiliation{%
  \institution{University College London}
  \city{London}
  \country{United Kingdom}
}
\email{emine.yilmaz@ucl.ac.uk}

\renewcommand{\shortauthors}{Lipani et al.}

\begin{abstract}
Information availability affects people's behavior and perception of the world. Notably, people rely on search engines to satisfy their need for information. 
Search engines deliver results relevant to user requests usually without being or making themselves accountable for the information they deliver, which may harm people's lives and, in turn, society.
This potential risk urges the development of evaluation mechanisms of bias in order to empower the user in judging the results of search engines. 
In this paper, we give a possible solution to measuring representation bias with respect to societal features for search engines and apply it to evaluating the gender representation bias for Google's Knowledge Graph Carousel for listing occupations.

\end{abstract}

%
%


\keywords{representation bias, rank bias measure, gender bias}

\maketitle


\section{Introduction}
%
Nowadays, search engines are increasingly used as primary tools to access information. The information that search engines point their users too often helps people take decisions that guide them through their lives, while it may also affect their judgement and perception of the world \cite{Kay:2015:URG:2702123.2702520}. Towards the end of the last century it has been observed that information systems can be biased \cite{Friedman:1996:BCS:230538.230561} and that biased information (i.e., information that does not represent the reality of the world people live in) can be harmful to how people relate to each-other and to how they evaluate their own decisions and opportunities \cite{Spencer1999}. 
The essential role that search engines have in influencing people's access to information urges the need to design evaluation mechanisms to assess search systems for potential biases that may negatively impact society.
This urge is further accentuated by the lack of transparency in how results to user searches are selected and displayed.
Such a bias assessment mechanism can be used as a diagnostic tool to monitor search systems results and make the systems more accountable, a desire that is raising to the attention of the research community \cite{roegiest-2019-facts-ir}.

Various measures to quantify bias in ranking have been previously proposed, e.g., \citet{yang2017measuring} define 3 measures starting from utility-based IR evaluation measures like RBP and nDCG; \citet{zehlike2017fa} define a statistical based method to evaluate and produce fair rankings. 
These metrics were then further discussed and improved by \citet{Gezici2021}.
In this short paper we focus on representation bias, also known as equality of opportunity \cite{NIPS2016_6374} which is rank-independent, while the mentioned works focus on statistical parity, i.e. a fair ranking system retrieves the same proportion of individuals across groups. 
S.~C.~Geyik et al \cite{geyik_fairness-aware_2019} define a ratio based metric for the feature representation at cut-off $k$ but all experiments further on are wrt. ranking and not rank-free representation bias.

Many countries enforce legislation to protect people from discrimination, be it at work or in the wider society. More concretely, such legislation aims to protect individuals from being discriminated as to societal features like outer aspect, education, sexual orientation, or geographical location. In this context, search engines bias evaluation can also be used by regulatory bodies in order to combat potential discrimination during search for information.

\begin{figure}[!t]
    \centering
    \resizebox{0.46\textwidth}{!}{
    \hspace{-1em}\includegraphics{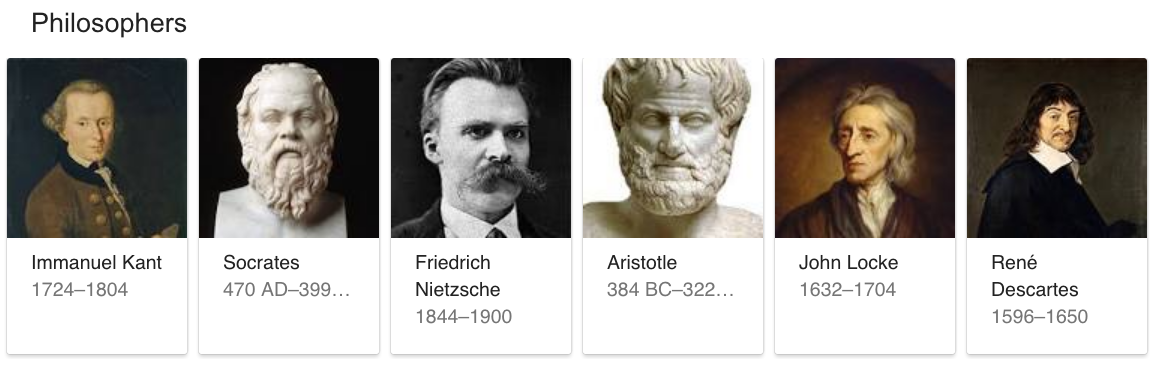}
    }
    \caption{
    Google's KGC as shown by Google when querying: `list of philosophers.'}
    \label{fig:list_of_philosophers}
\end{figure}

Traditionally search engines are evaluated in terms of effectiveness and efficiency. An effective system retrieves relevant information while an efficient one is a system that does it fast and with few resources.
We argue, here, that an orthogonal form of search engine evaluation is needed where the fair representation of relevant societal features in search results is assessed.
In other words, we argue for a measure to assess a search system's \emph{feature representation bias}. 
In this work we formalize the representation bias for categorical features and illustrate its use to analyze the gender representation bias in exponents of professions as displayed by the Knowledge Graph Carousel (KGC) featured by Google (Figure \ref{fig:list_of_philosophers}).
%
%

%
%
%

\section{Representation Bias}
\label{sec:bias}


The \emph{representation bias} of a search system is the preference of the system for or against a group of results relevant to a user query, that manifest a specific feature value $c \in \mathcal{C}$, with $\mathcal{C}$ a set of categorical features like age, education, gender. 
For a given feature $c \in \mathcal{C}$, we compute a system's representation bias with respect to this feature as the difference between the proportion of result documents manifesting the $c$ feature value (the \emph{model $c$-ratio}) and the proportion of all documents indexed by a search system that present the $c$ feature value (the \emph{target $c$-ratio}).

We give, now, the definitions and rationale behind the representation bias and its two components, model and target $c$-ratios. 
In this work, and for the sake of formalization simplicity, we consider $\mathcal{C}$ to be binary. The formalization can be generalized to any $\mathcal{C}$ size. 
\begin{definition}
Given the set of documents, $\mathcal{D}$, relevant to a user's input query, and a binary feature set, $\mathcal{C}$, we define the \textbf{\emph{target $c$-ratio}} as:
\begin{equation}%
    \mu_c = \frac{| \mathcal{D}^c | }{|\mathcal{D}|},
\end{equation}
where $\mathcal{D}^c$ is the set of documents in $\mathcal{D}$ that manifest feature $c \in \mathcal{C}$.
\end{definition}

Since search engines, however, show only a lower size $n$ subset $\mathcal{U}$ of the document set $\mathcal{D}$ found relevant to a query, we could define the \emph{target $c$-ratio at cut-off $n$} similarly, as:
\begin{equation}\label{eq:naive_estimator}
    \mu_c@n = \frac{| \mathcal{U}^c | }{n},
\end{equation}
with $n$ = $|\mathcal{U}|$ and $\mathcal{U}^c$ is the set of documents in $\mathcal{U}$ with feature $c$\footnote{Note that $\mu_c@n$ has values in the set $\{k/n; k,n \in \mathbb{N}, 0 \le n, 0 \leq k \leq n\}$.}. This definition is, though, insufficient explained below.

To measure a search system's representation bias our method compares the \emph{target $c$-ratio at cut-off n} with the \emph{model $c$-ratio} where the latter is defined as:
\begin{definition}%
Given the set of documents, $\mathcal{D}$, relevant to a user's input query, a binary feature set, $\mathcal{C}$, and a search result, $r = [d_1, \dots, d_m]$ with $d_1, \dots, d_m \in \mathcal{D}$ of documents ranked by the search system, 
we define the \textbf{\emph{model $c$-ratio at cut-off $n$ with respect to $r$}} as:
\begin{equation}
    \hat{\mu}^r_c@n = \frac{|\mathcal{D}_{r@n}^c|}{n},
\end{equation}
where $r@n$ is $r$ cut at rank $n$, $\mathcal{D}_{r@n}^c$ is the set of documents in $r@n$ that have the feature $c$.
\end{definition}

Ideally, $\mu_c$ and $\mu_c@n$ are the same, making a comparison between $\mu_c@n$ and $\hat{\mu}^r_c@n$, which we aim for with our method, straightforward. However, since 
$|\mathcal{U}| = n$ is usually much lower than $|\mathcal{D}|$, this is not always the case. Consider, for example, $\mu_c = 0.5$ and $n = 11$. Two result sets $\mathcal{D}_{r@n}$, one that contains 5 $c$-feature documents while the other displays 6 $c$-feature documents must be considered as equal from a representational bias point of view. Using the naive estimator in (\ref{eq:naive_estimator}), in certain circumstances, we would be measuring an non existing bias.
To this end, we define the unbiased version of the target $c$-ratio as follows:
\begin{definition}
For a given set of documents, $\mathcal{D}$, relevant to a user's input query, a binary feature set, $\mathcal{C}$, and a search result, $r$, the cut-off $n$, let 
$\delta_n = \mu_c \cdot n - \lfloor \mu_c \cdot n \rfloor$. 
We define the \textbf{\emph{target $c$-ratio at cut-off $n$ with respect to $r$}} as:
\begin{equation}\label{eq:t_n}
    {\mu_c^r@n} = 
    \frac{1}{n}
    \begin{cases}
    \lfloor \mu_c \cdot n \rfloor & \delta_n < 0.5 \\
    \argmin_{\{\lfloor \mu_c \cdot n \rfloor, \lceil \mu_c \cdot n \rceil\}}(|x - \hat{\mu}_c^r@n|) &  \delta_n = 0.5 \\
    \lceil \mu_c \cdot n \rceil & \delta_n > 0.5 \\
    \end{cases}.
\end{equation}
\end{definition}
\noindent The \emph{target $c$-ratio at cut-off $n$} here defined is the `ideal' model $c$-ratio, that is the observed $r$ is free of representational bias.
The way this is defined ensures that when computing the ideal $c$-ratio for a given $r$, in case of  ambiguity ($\delta_n = 0.5$) the solution given by the search engine is accepted as correct.

\begin{definition}
The \textbf{\emph{$c$-feature representation bias for a result set $r$}}  is the difference between the \emph{model} and the \emph{target} $c$-ratios with respect to $r$:
\begin{equation}\label{eq:b_n}
    {\beta}_c^r@n = \mu_c^r@n - \hat{\mu}_c^r@n 
\end{equation}
\end{definition}
\noindent This equation is defined in the set 
$\{k/n : k, n \in \mathbb{Z}, n > 0, |k| \leq n \}$.

When $\beta_c^r@n = 0$ we consider that $r$ is bias-free with respect to the feature $c$. 
When $\beta_c^r@n > 0$ the system result is biased towards $c$, with maximal bias when $\beta@n = 1$. 
When $\beta_c^r@n < 0$ the system result is biased against $c$, with maximal bias when $\beta@n = -1$. 

We can now compute the Mean representation Bias ($\text{MB}$) and its Standard deviation ($\text{SB}$) over a set of user queries (topics) for each feature value $c$. 
Note that when a search engine is biased towards one $c$ value for one topic and towards the other value for another topic, these two topics may cancel each other out when computing $\text{MB}$. To compensate for this effect we compute also the Mean of the Absolute Bias ($\text{MAB}$).

\begin{figure*}[!ht]
    \centering
    \resizebox{\textwidth}{!}{%
    \input{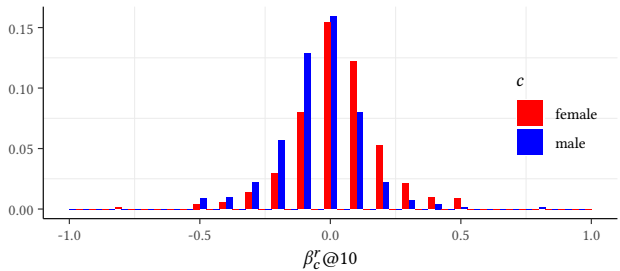}
    \input{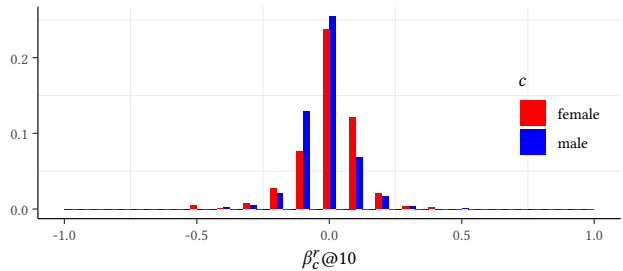}
    }
    \vspace{-2.5em}
    \caption{
    Gender representation bias histograms for $\mathcal{K}$ (left) and $\mathcal{R}$ (right)
    }
    \label{fig:histograms}
\end{figure*}

\section{Case Study}
\label{sec:case_study}

Using the feature ratios defined in the previous section, we measure the representation bias for gender (a binary feature) at cut-off $n = 10$. The user topics we consider are professions.
We measure the gender representation bias for Google's KGS. The code used to collect the dataset and ran the analysis is available at the following web-link: \url{https://github.com/aldolipani/TMASE}.
We note here that, as Google's results are profile dependent - to say the least - we have crawled the KGC for professions using a United States location and an incognito browsing mode.
Figure~\ref{fig:list_of_philosophers} shows an example of KGC display, located between the Google search bar and the results list. Such a KGC is displayed only when users give specific word patterns as input queries.

Though the list of word patterns that trigger the KGC display is not disclosed, based on our tests, we found the following triggering patterns: {\small\texttt{`list of [professions' noun]'}}, {\small\texttt{`top [professions' noun]'}}, {\small\texttt{`important [professions' noun]'}}.
In this analysis we focus only on the first pattern that triggers the KGC display, due to its (gender) neutral surface formulation. The connotative meaning of the {\small\texttt{`top'}} or {\small\texttt{`important'}} terms bring emotional and other associations we want to avoid.
In this context, the neutral connotations for queries like {\small\texttt{`list of [professions' noun]'}} demonstrate a \emph{neutral intent} to retrieve {\small\texttt{persons'}} associated with the {\small\texttt{professions'}}.
Although the search engine may return their results based on a predicted user preference, 
we believe that these queries in particular, due to their neutral intent, and based on the fact that Google does not make itself accountable by disclosing the feature used to rank these results (like `historical importance', `yearly income', etc.), 
should give results free of bias.

During our initial experiments, we found out that not all queries to list persons with a certain occupations trigger the KGC display. Therefore we extracted all professions available in Wikidata \cite{PellissierTanon:2016:FWG:2872427.2874809}, 
together with their gender feature annotations. This gave us a list of 3,374 professions for which we generated Google word patterns, and retained 454 professions which did trigger the KGC display. The analysis further on is done on these 474 professions. 
For cases where Google's KG gender annotation did not match Wikidata's, we did manual labelling.

As defined in the previous section, to calculate the target $c$-ratio for the results displayed in the KGC we first compute the \emph{target $c$-ratio}, $\mu_c$, for each profession. We select two document sets: (a) the full-length Google search result for professions, denoted by $\mathcal{R}$; and (b)
the Wikidata collaboratively edited knowledge base (KB) hosted by the Wikimedia Foundation, denoted by $\mathcal{K}$.

The results shown by KGC, as many online systems, change over time, so the following analysis is relative to data crawled on December 1st, 2019.

\begin{table}[!t]
    \centering
    \vspace{-1em}
    \caption{Summary of the data visualized in Fig.~\ref{fig:histograms}.}
    \resizebox{0.48\textwidth}{!}{%
    \begin{tabular}{r| >{\raggedleft\arraybackslash}p{1cm} | >{\centering\arraybackslash}p{1.cm} | >{\centering\arraybackslash}p{1.0cm} | >{\centering\arraybackslash}p{1.2cm} | >{\centering\arraybackslash}p{1cm} | >{\centering\arraybackslash}p{1cm}}
    \toprule
    & \rule{0pt}{2.5ex} $c$ & MB@10 & SB@10 & MAB@10 & $\min$ & $\max$\\ 
    \midrule
    \multirow{2}{*}{$\mathcal{K}$} 
    & female & -0.042 & 0.156 & 0.113 & -0.5 & 0.8\\
    & male   & \leavevmode\hphantom{-}0.026 & 0.170 & 0.121 & -0.8 & 0.5\\
    \hline
    \multirow{2}{*}{$\mathcal{R}$} 
    & female & -0.015 & 0.097 & 0.062 & -0.4 & 0.5\\
    & male   & \leavevmode\hphantom{-}0.001 & 0.113 & 0.072 & -0.5 & 0.4\\
    \bottomrule
    \end{tabular}
    }
    \label{tab:summary_hists}
    \vspace{-2em}
\end{table}

Figure~\ref{fig:histograms} shows the per gender histograms of KGC representation bias values in relation to $\mathcal{R}$ and $\mathcal{K}$ target c-ratios.
Considering the binary characteristic of gender we expected these histograms to be symmetric to each other, for each of the two cases.
However, these are not perfectly symmetric as for some queries the KGC has shown less than 10 results. Still, had the KGC been free of bias, we would now observe two histograms centered on zero with no spread. Since this is not the case we may conclude that the KGC is not free of gender representation bias. Additionally, the pair-wise asymmetry of the histograms indicates that when the search result is biased, it is more likely that the bias favours the male gender over the female one. 
Looking at the distribution bias wrt.~the $\mathcal{K}$ target gender-ratio we observe a stronger bias than the one wrt.~the $\mathcal{R}$ target gender-ration of the full search result set.
Moreover, we also observe that when we base our target distribution based on the KB we measure more bias than what we measure when the target distribution is calculated over the full search results.
Table~\ref{tab:summary_hists} summarizes the data visualized in Figure~\ref{fig:histograms}.

\begin{figure*}[!ht]
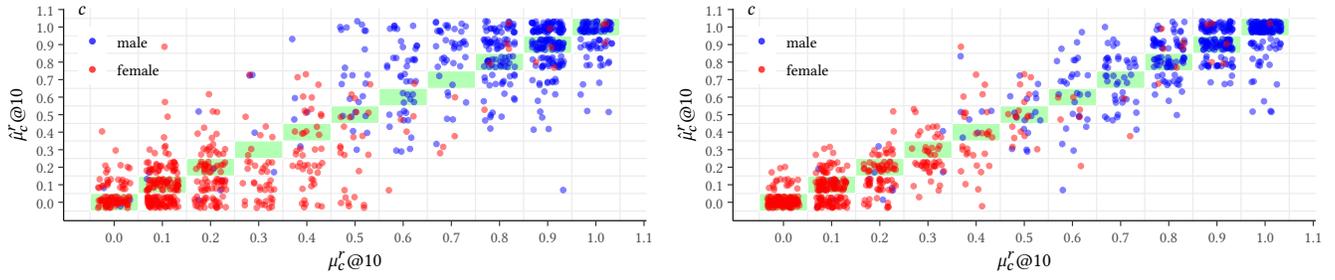

    \centering
    \resizebox{\textwidth}{!}{%
    \input{imgs/scatter_plot_gender_bias_kb}
    \input{imgs/scatter_plot_gender_bias_se}
    }
    \caption{
    Target vs. model gender-ratio for $\mathcal{K}$ (left) and $\mathcal{R}$ (right). 
    Every point represents a profession. A jitter is added to every data point, i.e.~points occurring in the same square of the grid have the same coordinates. The two sets of points for the male and female gender are symmetric. The points in the green squares are those free of bias. }
    \label{fig:scatter_plot}
\end{figure*}

Figure~\ref{fig:scatter_plot} compares the target (x-axis) and the model (y-axis) gender-ratios for each profession in $\mathcal{K}$ and $\mathcal{R}$. To every data point we added a controlled jitter that makes them fall within a square of the grid surrounding the original data point. Looking at the dispersion of points along the x-axis, we note that the male gender dominates many professions. The number of points per grid square decreases when moving towards the center, these are those professions equally represented by both genders. Only 4 professions are female-dominated. Would the KGC be free of bias, it would have placed every profession within the green squares, where the target and model gender-ratios are equal, i.e. bias is null. 

Tables~\ref{tab:summary_professions_kb} and \ref{tab:summary_professions_se} show the top 11 professions extracted from $\mathcal{K}$ and $\mathcal{R}$ for which the highest representation bias towards the male gender was observed, no gender-ratio bias was observed, and highest bias towards female gender was observed, respectively. 
The concrete values in these tables represent the female-ratios\footnote{Female- and male-ratio sum up to 1.}.
Table~\ref{tab:summary_professions_kb} identifies `announcer' as the profession over-represented by the male gender feature, and `archivists' as the profession over-represented by the female gender feature. 
For the $\mathcal{K}$ data `archivists' is over-represented by females, and `librarian` by the male feature.

From the data in these tables we observe that while the professions biased in favour of the male gender have a target gender-ratio of around 0.5, the target ratios for professions that are female gender-biased is around 0.1. This indicates that females, when fairly represented in the larger data ($\mathcal{R}$ or $\mathcal{K}$) are less represented in the KGC results, and when they are not represented in the larger data, their representation ratio in the KGC results is higher.

\begin{table*}[!ht]
    \centering
    \vspace{1em}
    \caption{Values are calculated with $c = \text{female}$. The column \emph{Unbiased} spans across every target ratio, from 0.0 to 1.0. 
    In this column, when more than one profession is found with the same target ratio, we selected the one with the largest population in $\mathcal{K}$. 
    The first and last columns are generated by ordering each profession based on its representation bias.}\label{tab:summary_professions_kb}
    \resizebox{\textwidth}{!}{%
    \begin{tabular}{r|c|c|c|r|c|c|c|r|c|c|c}
    \toprule
    \multicolumn{4}{c|}{\bfseries Biased Towards Males} & 
    \multicolumn{4}{c|}{\bfseries Unbiased} & 
    \multicolumn{4}{c}{\bfseries Biased Towards Females} \\ \midrule
    Profession & $\hat{\mu}^r_c@10$ & $\mu^r_c@10$ & $\beta^r_c@10$ & 
    Profession & $\hat{\mu}^r_c@10$ & $\mu^r_c@10$ & $\beta^r_c@10$ & 
    Profession & $\hat{\mu}^r_c@10$ & $\mu^r_c@10$ & $\beta^r_c@10$ \\ 
    \midrule
    announcer &0.0 &0.5 & -0.5 &              american football player &0.0 &0.0 &0.0 &    archivist &0.9 &0.1 &0.8 \\
    long jumper &0.0 &0.5 & -0.5 &            historian &0.1 &0.1 &0.0 &                   baker &0.6 &0.1 &0.5 \\
    high jumper &0.0 &0.5 & -0.5 &            songwriter &0.2 &0.2 &0.0 &                  school teacher &0.6 &0.2 &0.4 \\
    science writer &0.1 &0.6 & -0.5 &         illustrator &0.3 &0.3 &0.0 &                 modern pentathlete &0.5 &0.1 &0.4 \\
    rugby sevens player &0.0 &0.5 & -0.5 &    choreographer &0.4 &0.4 &0.0 &               church musician &0.4 &0.0 &0.4 \\
    cell biologist &0.0 &0.5 & -0.5 &         badminton player &0.5 &0.5 &0.0 &            drama teacher &0.7 &0.3 &0.4 \\
    clinical psychologist &0.0 &0.5 & -0.5 &  artistic gymnast &0.5 &0.5 &0.0 &            television presenter &0.7 &0.4 &0.3 \\
    piano teacher &0.0 &0.5 & -0.5 &          model &0.8 &0.8 &0.0 &                       scenographer &0.5 &0.2 &0.3 \\
    water polo player &0.0 &0.4 & -0.4 &      flight attendant &0.9 &0.9 &0.0 &            track cyclist &0.4 &0.1 &0.3 \\
    middle-distance runner &0.0 &0.4 & -0.4 & rhythmic gymnast & 1.0 & 1.0 & 0.0 &            skeleton racer &0.7 &0.4 &0.3 \\
    botanical illustrator &0.3 &0.7 & -0.4 &  glamour model & 1.0 & 1.0 &0.0 &               game author &0.3 &0.0 &0.3 \\
    \bottomrule
    \end{tabular}
    }
\end{table*}

\begin{table*}[!ht]
    \centering
    \caption{Values are arranged as in Table \ref{tab:summary_professions_kb} but 
    selected using $\mathcal{R}$. 
    }\label{tab:summary_professions_se}
    \resizebox{\textwidth}{!}{%
    \begin{tabular}{r|c|c|c|r|c|c|c|r|c|c|c}
    \toprule
    \multicolumn{4}{c|}{\bfseries Biased Towards Males} & 
    \multicolumn{4}{c|}{\bfseries Unbiased} & 
    \multicolumn{4}{c}{\bfseries Biased Towards Females} \\ \midrule
    Profession & $\hat{\mu}^r_c@10$ & $\mu^r_c@10$ & $\beta^r_c@10$ & 
    Profession & $\hat{\mu}^r_c@10$ & $\mu^r_c@10$ & $\beta^r_c@10$ & 
    Profession & $\hat{\mu}^r_c@10$ & $\mu^r_c@10$ & $\beta^r_c@10$\\ \midrule
    librarian &0.2 &0.6 & -0.4 &              officer of the french navy &0.0 &0.0 &0.0 &    archivist &0.9 &0.4 &0.5 \\
    draughts player &0.0 &0.4 & -0.4 &        war photographer &0.1 &0.1 &0.0 &              scenographer &0.5 &0.2 &0.3 \\
    long jumper &0.0 &0.3 & -0.3 &            table tennis player &0.2 &0.2 &0.0 &           video blogger &0.7 &0.4 &0.3 \\
    handball player &0.1 &0.4 & -0.3 &        fashion designer &0.3 &0.3 &0.0 &              drama teacher &0.7 &0.4 &0.3 \\
    translator &0.1 &0.4 & -0.3 &             alpine skier &0.4 &0.4 &0.0 &                  sailor &0.2 &0.0 &0.2 \\
    classical archaeologist &0.0 &0.2 & -0.2 &       vj &0.5 &0.5 &0.0 &                            lighting designer &0.3 &0.1 &0.2 \\
    high jumper &0.0 &0.2 & -0.2 &     sex educator &0.6 &0.6 &0.0 &                  chemist &0.3 &0.1 &0.2 \\
    talk show host &0.0 &0.2 & -0.2 &                beach volleyball player &0.6 &0.6 &0.0 &       fighter pilot &0.2 &0.0 &0.2 \\
    sound artist &0.0 &0.2 & -0.2 &            softball player &0.8 &0.8 &0.0 &               musical theatre actor &0.6 &0.4 &0.2 \\
    executive &0.0 &0.2 & -0.2 &              domestic worker &0.9 &0.9 &0.0 &               television presenter &0.7 &0.5 &0.2 \\
    science journalist &0.1 &0.3 & -0.2 &    ballerina & 1.0 & 1.0 &0.0 &                   skeleton racer &0.7 &0.5 &0.2 \\
    \bottomrule
    \end{tabular}
    }
\end{table*}

\section{Conclusion}
\label{sec:conclusion}

We have defined concepts to help evaluate the representation bias with respect to specific societal features in online search engines. Such evaluations, if provided together with search results, will allow users to assess potential representation biases.
In a concrete case study, we have used the defined concepts to assess the representation bias of the Google's KGC with respect to gender and professions. Assuming the correctness of the population frequency of the documents available to KGC, we conclude that it suffer from gender representation bias.

A key aspect of the method introduced in this paper is the assumption of complete knowledge of the population frequency for each feature value. In our case study, the \emph{feature ratios} have been calculated based on data that the search system provider made available and on a collaboratively knowledge base (Wikidata). 
However, better statistics for $c$-ratios could be aimed for, e.g.~official statistics published by government agencies.

Finally, we have shown how such a representation bias evaluation can be performed using a concrete case study. In this case study we have analyzed the gender representation bias of the Google's KGC, and shown that, assuming the correctness of the population means, it suffers from gender representation bias.
%
We must be aware, though, that the systems' selection of documents to be displayed (by KGC, for example), is most of the time not transparent, and other features not openly accounted for, like 'historical relevance' or 'yearly income' may affect the final result shown to a user. 
At the same time we must distinguish, between detecting the source of bias and measuring a representation bias. The work described in this paper may help a user hypothesize on sources of bias once they have been detected by measuring the representation bias.

\bibliographystyle{ACM-Reference-Format}
\bibliography{main}


\begin{thebibliography}{00}


\ifx \showCODEN    \undefined \def \showCODEN     #1{\unskip}     \fi
\ifx \showDOI      \undefined \def \showDOI       #1{#1}\fi
\ifx \showISBNx    \undefined \def \showISBNx     #1{\unskip}     \fi
\ifx \showISBNxiii \undefined \def \showISBNxiii  #1{\unskip}     \fi
\ifx \showISSN     \undefined \def \showISSN      #1{\unskip}     \fi
\ifx \showLCCN     \undefined \def \showLCCN      #1{\unskip}     \fi
\ifx \shownote     \undefined \def \shownote      #1{#1}          \fi
\ifx \showarticletitle \undefined \def \showarticletitle #1{#1}   \fi
\ifx \showURL      \undefined \def \showURL       {\relax}        \fi
\providecommand\bibfield[2]{#2}
\providecommand\bibinfo[2]{#2}
\providecommand\natexlab[1]{#1}
\providecommand\showeprint[2][]{arXiv:#2}

\bibitem[\protect\citeauthoryear{et~al.}{et~al.}{2019a}]%
        {roegiest-2019-facts-ir}
\bibfield{author}{\bibinfo{person}{Adam~Roegiest et al.}}
  \bibinfo{year}{2019}\natexlab{a}.
\newblock \showarticletitle{FACTS-IR: Fairness, Accountability,
  Confidentiality, Transparency, and Safety in Information Retrieval}.
\newblock \bibinfo{journal}{{\em SIGIR Forum\/}} \bibinfo{volume}{53},
  \bibinfo{number}{2} (\bibinfo{year}{2019}).
\newblock


\bibitem[\protect\citeauthoryear{et~al.}{et~al.}{2016}]%
        {PellissierTanon:2016:FWG:2872427.2874809}
\bibfield{author}{\bibinfo{person}{Pellissier~Tanon et al.}}
  \bibinfo{year}{2016}\natexlab{}.
\newblock \showarticletitle{From Freebase to Wikidata: The Great Migration}. In
  \bibinfo{booktitle}{{\em Proc.~of WWW '16}}.
\newblock


\bibitem[\protect\citeauthoryear{et~al.}{et~al.}{2019b}]%
        {geyik_fairness-aware_2019}
\bibfield{author}{\bibinfo{person}{S.~C.~Geyik et al.}}
  \bibinfo{year}{2019}\natexlab{b}.
\newblock \showarticletitle{Fairness-{Aware} {Ranking} in {Search} \&
  {Recommendation} {Systems} with {Application} to {LinkedIn} {Talent}
  {Search}}. In \bibinfo{booktitle}{{\em Proc. of {KDD}'19}}.
\newblock
\showISBNx{978-1-4503-6201-6}


\bibitem[\protect\citeauthoryear{Friedman and Nissenbaum}{Friedman and
  Nissenbaum}{1996}]%
        {Friedman:1996:BCS:230538.230561}
\bibfield{author}{\bibinfo{person}{Batya Friedman} {and} \bibinfo{person}{Helen
  Nissenbaum}.} \bibinfo{year}{1996}\natexlab{}.
\newblock \showarticletitle{Bias in Computer Systems}.
\newblock \bibinfo{journal}{{\em ACM Trans. Inf. Syst.\/}}
  \bibinfo{volume}{14}, \bibinfo{number}{3} (\bibinfo{date}{July}
  \bibinfo{year}{1996}), \bibinfo{pages}{330--347}.
\newblock
\showISSN{1046-8188}


\bibitem[\protect\citeauthoryear{Gezici, Lipani, Saygin, and Yilmaz}{Gezici
  et~al\mbox{.}}{2021}]%
        {Gezici2021}
\bibfield{author}{\bibinfo{person}{Gizem Gezici}, \bibinfo{person}{Aldo
  Lipani}, \bibinfo{person}{Yucel Saygin}, {and} \bibinfo{person}{Emine
  Yilmaz}.} \bibinfo{year}{2021}\natexlab{}.
\newblock \showarticletitle{{Evaluation metrics for measuring bias in search
  engine results}}.
\newblock \bibinfo{journal}{{\em Information Retrieval Journal\/}}
  \bibinfo{volume}{24}, \bibinfo{number}{2} (\bibinfo{year}{2021}).
\newblock


\bibitem[\protect\citeauthoryear{Hardt, Price, and Srebro}{Hardt
  et~al\mbox{.}}{2016}]%
        {NIPS2016_6374}
\bibfield{author}{\bibinfo{person}{Moritz Hardt}, \bibinfo{person}{Eric Price},
  {and} \bibinfo{person}{Nati Srebro}.} \bibinfo{year}{2016}\natexlab{}.
\newblock \showarticletitle{Equality of Opportunity in Supervised Learning}. In
  \bibinfo{booktitle}{{\em Proc.~of NeuIPS '16}}.
\newblock


\bibitem[\protect\citeauthoryear{Kay, Matuszek, and Munson}{Kay
  et~al\mbox{.}}{2015}]%
        {Kay:2015:URG:2702123.2702520}
\bibfield{author}{\bibinfo{person}{Matthew Kay}, \bibinfo{person}{Cynthia
  Matuszek}, {and} \bibinfo{person}{Sean~A. Munson}.}
  \bibinfo{year}{2015}\natexlab{}.
\newblock \showarticletitle{Unequal Repr.~and Gender Stereotypes in Image
  Search Results for Occupations}. In \bibinfo{booktitle}{{\em Proc.~of CHI
  '15}}.
\newblock


\bibitem[\protect\citeauthoryear{Spencer, Steele, and Quinn}{Spencer
  et~al\mbox{.}}{1999}]%
        {Spencer1999}
\bibfield{author}{\bibinfo{person}{Steven~J. Spencer},
  \bibinfo{person}{Claude~M. Steele}, {and} \bibinfo{person}{Diane~M. Quinn}.}
  \bibinfo{year}{1999}\natexlab{}.
\newblock \showarticletitle{{Stereotype Threat and Women's Math Performance}}.
\newblock \bibinfo{journal}{{\em Journal of Experimental Social Psych.\/}}
  \bibinfo{volume}{35} (\bibinfo{year}{1999}).
\newblock
\showISSN{0022-1031}


\bibitem[\protect\citeauthoryear{Yang and Stoyanovich}{Yang and
  Stoyanovich}{2017}]%
        {yang2017measuring}
\bibfield{author}{\bibinfo{person}{Ke Yang} {and} \bibinfo{person}{Julia
  Stoyanovich}.} \bibinfo{year}{2017}\natexlab{}.
\newblock \showarticletitle{Measuring fairness in ranked outputs}. In
  \bibinfo{booktitle}{{\em Proc.~of SSDBM '17}}.
\newblock


\bibitem[\protect\citeauthoryear{Zehlike~et al.}{Zehlike~et al.}{2017}]%
        {zehlike2017fa}
\bibfield{author}{\bibinfo{person}{M. Zehlike~et al.}}
  \bibinfo{year}{2017}\natexlab{}.
\newblock \showarticletitle{{FA*IR}: A fair top-k ranking algorithm}. In
  \bibinfo{booktitle}{{\em Proc.~CIKM '17}}.
\newblock


\end{thebibliography}

\end{document}